\documentclass[letterpaper, 10 pt, conference]{ieeeconf}
\IEEEoverridecommandlockouts
\let\labelindent\relax
\usepackage{enumitem}

\usepackage{cite} 		  
\usepackage{xcolor}       
\usepackage{hyperref}	  
\usepackage{xurl}
\hypersetup{colorlinks=true,linkcolor=blue,citecolor=blue,breaklinks={true}} 
\usepackage[normalem]{ulem}

\setlist[itemize]{leftmargin=*}  

\usepackage{graphicx}
\usepackage{sidecap}
\usepackage[font=footnotesize,labelfont=bf,justification=justified]{caption}
\graphicspath{{./FIG/}}
\DeclareGraphicsExtensions{.eps,.png,.jpg,.jpeg,.bmp,.gif,.pdf}

\usepackage{tabularx,booktabs}	
\usepackage{multirow}	

\usepackage{algorithm}
\usepackage{algorithmic}

\usepackage{amsmath}      		
\usepackage{amssymb}	  		
\usepackage{bm}		      		
\usepackage{physics}	  		
\usepackage{mathtools}			
\usepackage{dsfont}		  		
\usepackage{soul}				
\usepackage{relsize}			

\newcommand{\R}{\mathbb{R}}		
\newcommand{\transp}{\mathsf{T}}					

\usepackage{amsthm}

\newtheorem{thm}{Theorem}

\newtheorem{lem}{Lemma}
\theoremstyle{definition}

\newtheorem{rem}{Remark}

\newcommand{\skipsum}[1]{%
  \mathop{{\vphantom{\sum}}_{#1}\kern-\scriptspace}\!\sum\nolimits
}

\title{\LARGE \bf
Distributed Lyapunov Functions for Nonlinear Networks
}

\author{Yiming Wang, Arthur N. Montanari, and Adilson E. Motter, \textit{Senior Member, IEEE}
\thanks{This work was supported by the U.S. Army Research Office MURI under Grant W911NF-24-1-0228. Y. Wang, A. N. Montanari, and A. E. Motter are with the Center for Network Dynamics and the Department of Physics \& Astronomy, Northwestern University, Evanston, IL 60208, USA (e-mails: yiming.wang2@northwestern.edu, arthur.montanari@northwestern.edu, motter@northwestern.edu).
A. E. Motter is also with the Department of Engineering Sciences \& Applied Mathematics and the Northwestern Institute on Complex Systems, Northwestern University, Evanston, IL 60208, USA.}
\vspace{-0.3cm}
}

\begin{document}


\maketitle
\thispagestyle{empty} 

\begin{abstract}
Nonlinear networks are often multistable, exhibiting coexisting stable states with competing regions of attraction (ROAs). As a result, ROAs can have complex ``tentacle-like'' morphologies that are challenging to characterize analytically or computationally. In addition, the high dimensionality of the state space prohibits the automated construction of Lyapunov functions using state-of-the-art optimization methods, such as sum-of-squares (SOS) programming. In this letter, we propose a distributed approach for the construction of Lyapunov functions based solely on local information. To this end, we establish an augmented comparison lemma that characterizes the existence conditions of partial Lyapunov functions, while also accounting for residual effects caused by the associated dimensionality reduction. These theoretical results allow us to formulate an SOS optimization that iteratively constructs such partial functions, whose aggregation forms a composite Lyapunov function. The resulting composite function provides accurate convex approximations of both the volumes and shapes of the ROAs. We validate our method on networks of van der Pol and Ising oscillators, demonstrating its effectiveness in characterizing high-dimensional systems with non-convex ROAs.

\end{abstract}

\smallskip
Published in \textit{IEEE Control Systems Letters}, 9:486-491 (2025), DOI: \href{https://doi.org/10.1109/LCSYS.2025.3573881}{10.1109/LCSYS.2025.3573881}. 

\section{Introduction}
\label{sec:introduction}
Multistability is a common property of large  networks of coupled dynamical systems \cite{wiley2006size,yiming1}. These systems are characterized by the coexistence of attractors, including stable equilibrium points (SEPs), which partition the state space into distinct regions of attraction (ROAs). Due to nonlinearity and high dimensionality, such ROAs can exhibit highly non-convex morphologies, often resembling ``octopus-like'' shapes  with elongated ``tentacles'' along specific directions of the state space \cite{zhang2021basins}. Characterizing the ROAs of multistable networks is important for the theory of oscillator networks \cite{wiley2006size,cheng2024control} and applications ranging from power-grid stability analysis \cite{yiming1} to data-driven control \cite{guo2021data}.

Highly accurate estimation of ROAs can be achieved for two- and three-dimensional systems \cite{packard2010help,anghel2013algorithmic}, yet scaling such results to large-scale networks remains challenging. 
Direct numerical evolution of the system for ROA estimation becomes computationally prohibitive as the required number of sampled initial conditions grows exponentially with state-space dimension. Scalable approaches that sample ROAs on low-dimensional sections of the state space offer insights\cite{menck2013basin} but yield limited information about the stability margins in the full state space. 
In contrast, Lyapunov functions provide a rigorous framework for estimating stability margins. However, most existing approaches to construct them tend to yield conservative ROA estimates\cite{khalil2002nonlinear}. 
For instance, quadratic Lyapunov functions, while analytically and computationally tractable for large systems \cite{cheng2003quadratic,cheng2024control}, constrain the ROAs to ellipsoids. Monte Carlo sampling, eigenvalue methods, and passivity theory have all been explored to refine these bounds \cite{yiming2}, and yet they still fail to capture the intricate geometry of ROAs commonly found in network systems.

To overcome these limitations, recent advances in the construction of Lyapunov functions explored model-based physical principles (e.g., energy functions in mechanical and electrical systems)  \cite{zhang2016transient} or data-based methods \cite{lee2024note}. While effective in certain domains, these approaches rely on expert-designed models or extensive training data, which may not generalize across different parameter settings. 
Recent work has combined elements of both approaches, integrating machine learning with satisfiability modulo theories for automated Lyapunov function synthesis \cite{abate2020formal,liu2025physics}. These hybrid methods show potential in capturing complex dynamics, as demonstrated for various oscillator networks. However, their centralized formulation poses scalability challenges when applied to large-scale systems. Ref. \cite{jena2021distributed} proposes a scalable alternative by enforcing local dissipativity conditions. However, this approach requires linearization of the network couplings, which typically introduces conservatism.

In contrast with the approaches above, SOS-based methods provide a convex optimization framework for synthesizing Lyapunov functions, with dimensional lifting techniques enabling its possible application even in systems described by non-polynomial equations.
Although these methods can provide accurate ROA estimation and be in principle applied to a wide range of systems \cite{papachristodoulou2002construction,anghel2013algorithmic}, currently they are not computationally scalable to systems with more than a few variables.
To reduce this burden, prior work has applied the comparison lemma to synthesize \textit{partial} Lyapunov functions, which are valid only on low-dimensional subspaces \cite{kundu2017multiple}. Yet, no scalable method has been developed for the construction of \textit{composite} Lyapunov functions that provide accurate ROA estimation in the full state space of high-dimensional systems. Consequently, previous research has been unable to systematically characterize the complex morphology of high-dimensional ROAs. 

In this letter, we propose a distributed approach to interactively construct Lyapunov functions based on local information. This is achieved by establishing an augmented comparison lemma for the existence conditions of partial Lyapunov functions.
The distributed approach is thus formulated using a set of SOS optimization problems for the construction of partial functions, which can be solved in parallel and then aggregated into a composite Lyapunov function. Each SOS optimization is solved iteratively in order to optimize both the volume and shape of the ROA estimate.
We apply our method to networks of van der Pol and Ising oscillators, demonstrating its effectiveness in capturing the dependence of the ROAs' morphology on the system parameters.

\section{Problem Formulation}

Consider a network of $N$ coupled dynamical systems,
\begin{equation}
\label{eq.nonlinearsys}
    \dot{\bm x}_{i} = \bm f_i(\bm x_i) + \sum_{j=1}^N K_{ij}\bm g(\bm x_i,\bm x_j), \quad i = 1,\ldots,N,
\end{equation}

\vspace{-0.1cm}
\noindent 
where $\bm x_i = (x_{i,1}, \ldots, x_{i,m})^\transp \in \mathcal S^m \subseteq \R^m$ is the state vector of node $i$ and $K\in\R^{N\times N}$ is the adjacency matrix of the underlying network. The functions $\bm f_i \, : \, \mathcal S^m \mapsto \mathcal S^m$ and $\bm g_i \, : \, \mathcal S^{2m} \mapsto \mathcal S^m$ are smooth nonlinear mappings that characterize the nodal dynamics and pairwise coupling, respectively. The full-state vector is given by $\bm x = (\bm x_1, \dots, \bm x_N)^\transp \in \mathcal S^n\subseteq\R^n$, where $n = Nm$ is the state-space dimension.

Let \( \mathcal D \subseteq \mathcal S^n \) be an open set containing the equilibrium point \( \bm x^* = 0 \). 
We define a Lyapunov function \( V(\bm x): \mathcal D \mapsto \mathbb{R} \) as a differentiable function satisfying the conditions: 
1) \( V(0) = 0 \), 
2) \( V(\bm x) > 0 \), \(\forall \bm x \in \mathcal D \backslash \{0\} \), and 
3) \( \dot{V}(\bm x) \leq 0 \), \( \forall \bm x \in \mathcal D \backslash \{0\} \).
Furthermore, for some constant \( \eta > 0 \), there exists a positively invariant set \( \Omega = \{ \bm x \in \mathcal D : V(\bm x) \leq \eta \} \), which provides an \textit{estimate} of the ROA of the equilibrium $\bm x^*$ based on the Lyapunov function. The equilibrium \( \bm x^* = 0 \) is exponentially stable if there exists a decay rate \( \lambda > 0 \) so that
$\dot{V}{(t)} \leq -\lambda V(\bm x)$, $\forall \bm x \in \mathcal D$  \cite[Th. 4.10]{khalil2002nonlinear}.   
The following comparison lemma provides an upper bound for continuous functions, which is used in our distributed approach.

\begin{lem} \textnormal{\cite[Lemma 3.4]{khalil2002nonlinear}}
Consider the scalar differential equation $\dot{W} = h(t, W)$, where $h \, : \, \mathcal D \mapsto \R$ is a smooth and continuous function for all $t\geq 0$. Let $V(t) \, : \, \mathcal D \mapsto \R$ be a continuous function that satisfies the differential inequality $\dot V(t) \leq h(t,W(t))$, for $V(0)\leq W(0)$. Then, $V(t)\leq W(t)$.
\label{lem.classicalcomparison}
\end{lem}
\vspace{-0.3cm}

An automated approach for the construction of a Lyapunov function relies on SOS polynomials of the form

\vspace{-0.3cm}
\begin{equation}
\label{eq.sospolynomial}
    V(\bm x) = \sum_{i\in\mathcal I} \Big(\sum_{j=1}^{P} c_{ij} m_i(\bm x)\Big)^2,
\end{equation}
\vspace{-0.2cm}

\noindent
where $m_i(\bm x)$ denotes a monomial of degree $i$ and $\mathcal I \subseteq \{0,1,\ldots,D\}$ specifies the set of monomial degrees used in the polynomial, with maximum degree $D$ and $P=\binom{n}{i}$. 
Finding all coefficients $c_{ij}> 0$ so that $V$ satisfies the Lyapunov function conditions can, in principle, be achieved through SOS optimization. However, the number of monomial terms in $V$ grows exponentially with $n$, rendering direct SOS optimization computationally prohibitive for large-scale systems. 

To address this challenge, we propose a \textit{distributed} approach for constructing Lyapunov functions by decomposing the high-dimensional optimization problem of finding $V$ into a set of constrained low-complexity problems that can be solved in parallel. We formalize this approach by defining  the following \textit{constrained subspaces}:
\begin{equation}
    \mathcal S_{p} = \{ \bm x\in\mathcal S^n \, : \, \bm x_j = 0, \,\, \forall j \notin \mathcal V_{p} \}, \,\,\,\, p = 1,\ldots,L,
\end{equation}
\noindent where $\mathcal V_{p} \subseteq \{1,\ldots,N\}$ is a subset of nodes with cardinality $ 1\leq |\mathcal V_{p}|=k \leq N$ and $L=\binom{N}{k}$ is the number of distinct subsets. For each subset $\mathcal V_{p}$, we define the projection matrix $P_{p}\in\R^{N\times N}$ as a diagonal matrix with entry $(P_{p})_{jj} = 1$ if $j\in\mathcal V_{p}$ and $0$ otherwise. It follows that $(P_{p}\otimes I_m)\bm x\in\mathcal S_{p}$ for all $\bm x\in\mathcal S^n$, where $I_m$ is an identity matrix of size $m$.

Now, suppose there exists a \textit{composite} Lyapunov function $V_{\rm c} \, : \, \mathcal D \mapsto \R$. For each node subset $\mathcal V_{p}$, let $V_{p} \, : \, \mathcal D_{p} \mapsto \R$ be a \textit{partial} function defined in the subset $\mathcal D_{p}\subseteq\mathcal S_{p}$. We can then define the following residual function:
\begin{equation}
    R_{p}(\bm x) = V_{\rm c}(\bm x) - V_{p}\left( (P_{p}\otimes I_m)\bm x\right).
\label{eq.Vprojection}
\end{equation}

\begin{figure}
    \centering
    \includegraphics[width=0.8\columnwidth]{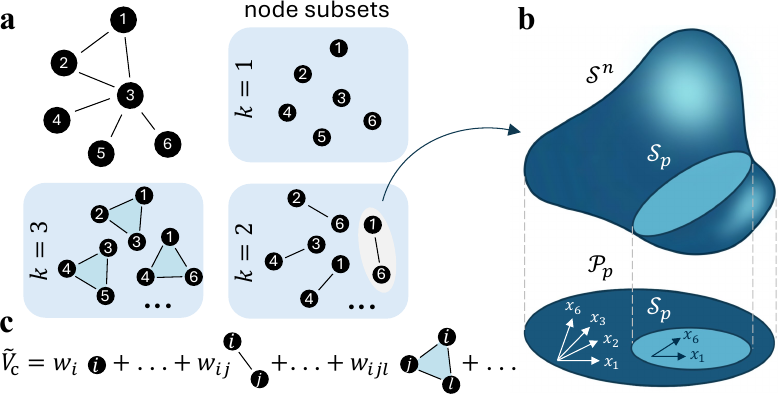}
    \caption{Distributed construction of a composite Lyapunov function. (\textbf{a}) Node subsets considered in the partial Lyapunov functions $V_{p}$ of a 6-node network for set cardinalities $k=1,2,3$. (\textbf{b}) Schematics of the constrained subspaces $\mathcal S_{p}$ and $\mathcal P_{p}$ for a representative function $V_{p}$ (where $\mathcal V_{p}=\{1,6\}$), highlighting the unconstrained variables in these subspaces. The subspaces $\mathcal P_{p}$ are defined in  Remark~\ref{rem.pkl}. (\textbf{c}) Composite Lyapunov function obtained by a linear combination of all partial functions.}
    \label{fig.diagram} 
    \vspace{-15pt}
\end{figure}

\noindent
In what follows, we establish the conditions under which all partial functions \( V_{p} \) serve as Lyapunov functions, and propose an SOS optimization framework to identify them. By combining these partial functions, we obtain a \textit{distributed approximation} of the composite Lyapunov function, \( \widetilde{V}_{\rm c} : \mathcal{D} \mapsto \mathbb{R} \), that estimates the ROA of an SEP $\bm x^*$ by maximizing the associated stability margin $\eta$ while ensuring that $\widetilde V_{\rm c}$ satisfies the Lyapunov function conditions. In particular, we provide a theoretical framework that explicitly accounts for the residual term in Eq.~\eqref{eq.Vprojection} in order to mitigate the conservativeness of the partial Lyapunov functions.

\begin{rem}
   Fig.~\ref{fig.diagram} summarizes our distributed approach. 
   For $N=6$, $m=1$, $k=2$, and $\mathcal{V}_{p} = \{1,6\}$, the projection matrix is defined as $P_{p} = \operatorname{diag}(1,0,0,0,0,1)$.
   Hence, the subspace $\mathcal S_{p}$ contains all $n$-dimensional vectors of form $(\bm x_1,0,0,0,0,\bm x_6)^\transp$. Note that $\bm x^*\in\mathcal D_{p}$ since the SEP is shifted to the origin $\bm x^*=0$.
   For a composite Lyapunov function with SOS form $V_{\rm c} = \sum_{i=1}^4 x_i^2 + (x_5 + x_6)^2$ and a quadratic partial function $V_{p} = x_1^2 + x_6^2$, it follows from Eq.~\eqref{eq.Vprojection} that $R_{p} = \sum_{i=2}^5 x_i^2 + 2x_5 x_6$.
\end{rem}

\section{Distributed Lyapunov Functions}

The residual function \( R_{p} \) has a crucial role in the distributed construction of $\widetilde V_{\rm c}$. 
Since the domain of \( R_{p} \) does not necessarily contain the SEP, Lemma~\ref{lem.classicalcomparison} cannot be applied to prove exponential stability.
Furthermore, neglecting \( R_{p} \) introduces non-conservative errors in the construction of the partial function $V_{p}$, which compromises estimation.
We thus propose the following \textit{augmented} comparison lemma, which directly accounts for all residual functions \( R_{p} \).

\vspace{-2pt}
\begin{lem}
\label{lem.augmentedcomparison}
Let $\bm V(t) = (V_{1},\ldots,V_{L})^\transp$ and $\bm R(t) = (R_{1},\ldots,R_{L})^\transp$ be column vectors so that $V_{p}$ and $R_{p}$ satisfy Eq.~\eqref{eq.Vprojection} for some $1\leq k\leq N$. 
Consider the element-wise differential inequality
\begin{equation}
\label{eq.differentialineq}
    \dot{\bm V}{(t)} \leq A \bm V(t) + B \bm R(t), \quad A\bm V(0)+B\bm R(0)=\bm c,
\end{equation}
\noindent
where $A,B\in\R^{L \times L}$ and $\bm c\in\R^{L}$.
If $A-B$ is a Metzler matrix, then the system  $\dot{\bm U}{(t)} = A \bm U(t) + B \bm R(t)$ provides the following upper bound:
\begin{equation}
\label{eq.contraction}
    \bm V(0) = \bm U(0) \Rightarrow \bm V(t) \leq \bm U(t), \quad \forall t \geq 0.
\end{equation}
\end{lem}
\vspace{-0.6cm}

\hspace{-8.15cm}
\begin{proof}
From Eq.~\eqref{eq.Vprojection}, it follows that $\bm R(t) = V_{\text{c}}(t)\mathbf{1}_{L} - \bm V(t)$, where $\mathbf{1}_L$ denotes an $L$-dimensional vector of ones. Substituting it into inequality \eqref{eq.differentialineq} yields $\dot{\bm V}{(t)} \leq (A - B) \bm V(t) + B  V_{\rm c}(t)\mathbf{1}_{L}$. Define the error function $\bm E(t)= \bm V(t) - \bm U(t)$. Taking its time derivative, we obtain
\begin{equation}
    \dot{\bm E}{(t)} = \dot{\bm V}{(t)} - \dot{\bm U}{(t)} \leq (A - B)(\bm V(t) - \bm U(t)).
\end{equation}
\noindent
If \( A - B \) is a Metzler matrix, then it follows from Ref.~\cite[p. 32]{bellman1962vector} that \( \bm E(t) \leq 0 \) for all \( t \geq 0 \), which leads to Eq.~\eqref{eq.contraction}.
\QED
\end{proof}

By ensuring that the conditions of Lemma~\ref{lem.augmentedcomparison} on matrices \( A \) and \( B \) are met, we establish that all partial functions $V_{p}$ are upper bounded by Eq.~\eqref{eq.contraction} and that\textemdash under further constraints\textemdash their time derivatives are also bounded as follows.

\vspace{-3pt}
\begin{thm}
\label{thm.locallyap}
Suppose the equilibrium $\bm x^* \in \mathcal D$ is stable, there exists a composite Lyapunov function $ V_{\rm c} : \mathcal D\mapsto\R$ , and the following conditions hold for each $p=1,\ldots,L$:
\begin{enumerate}[label=\textit{(\roman*)}, align=left, leftmargin=*, nosep]
    \item $A_{p r} - B_{p r} > 0$ for all $r=1,\ldots,N$ and $r \neq p $,
    \item $\sum_{r=1}^{L}  (A_{p r} - B_{p r}) \leq 0 $,
    \item  $\sum_{r=1}^{L}  B_{p r} \geq 0$,
\end{enumerate}
\noindent
where $A_{p r}$ and $B_{p r}$ denote the $(p, r)$th elements of matrices $A$ and $B$, respectively. Then, for all $p$, there exists functions $V_{p}$ satisfying
\vspace{-0.1cm}
\begin{equation}
    \dot{V}_{{p}} \leq \sum_{r=1}^{L} \Big[ \underbrace{( A_{p r}-B_{p r}) V_{r}  + B_{p r} V_{\rm {c}}}_{\Theta_{p, r}} \Big].
\label{eq.lem.timederivative}
\end{equation}
\end{thm}
\vspace{-0.3cm}

\vspace{-1pt}
\hspace{-1.2cm}
\begin{proof}
Note that condition (i), for all $p = 1,\ldots,L$, 
 is equivalent to requiring that \( A - B \)  be a Metzler matrix. Thus, if conditions (i)--(ii) hold, it follows from Lemma ~\ref{lem.augmentedcomparison} that \( \dot{V}_{{p}} \leq \sum_{r=1}^{L} ( A_{p r}-B_{p r}) V_{r} \leq 0 \), $\forall p$. Since $\dot V_{\rm c}\leq 0$, and condition (iii) is satisfied, it follows that \( \dot{V}_{\text{c}} \leq \sum_{r=1}^{L} B_{p r} V_{\text{c}} \). By combining these observations, we obtain Eq.~\eqref{eq.lem.timederivative}. \QED\end{proof}

\vspace{-2pt}
\begin{rem}
    If $V_{p}$ is described by an SOS polynomial, then $V_{p}(\bm x^*) = 0$ and $V_{p}(\bm x)>0$, $\forall \bm x\in\mathcal D_{p}\backslash\{0\}$. In addition, if Eq.~\eqref{eq.lem.timederivative} is satisfied, then $\dot V_{p}(\bm x) \leq 0$, $\forall p$. Thus, each function $V_{p}$ qualifies as a \textit{partial} Lyapunov function.
    Notably, each partial function \( V_{p} \) can be computed through SOS optimization \textit{in parallel} for each subset $\mathcal V_{p}$, given that Eq.~\eqref{eq.lem.timederivative} depends only on the \( p \)th rows of \( A \) and \( B \).
\end{rem}

\subsection{Optimal Estimation of the Partial Regions of Attraction}
Building on Theorem \ref{thm.locallyap}, we formulate an SOS optimization problem to construct partial Lyapunov functions.
Assume there exists a scalar \( \nu_{p} \geq 0 \) so that \( \gamma_{\text{c}} - V_{\text{c}}(\bm x) \geq 0 \) holds for all \( \bm x \in \{ \mathcal D_{p} \, : \, \gamma_{p} - V_{p}(\bm x) - \nu_{p}\bm x^\transp \bm x \geq 0 \} \backslash \{0\} \). 
It follows that any subset \( \Omega_{p} = \{ \bm x \in \mathcal{D}_{p} : V_{p}(\bm x) \leq \gamma_{p} - \nu_p \bm x^\transp \bm x\} \) is positively invariant and lies within the whole ROA \( \Omega \).
Our goal is to maximize the volume of $\Omega_{p}$ for each function $V_{p}$, which is equivalent to maximizing $\gamma_{p}$, as in
\vspace{-0.2cm}
\begin{subequations}
\label{eq.sosoptimization}
\begin{align}
    &\hspace{-0.45cm} \max_{\Xi} \,\,  \,\, \gamma_{p}     \label{eq:9a} \\
    \hspace{-0.2cm}\text{s.t.} &\,\, 
    s_{p} (V_{p} - \gamma_{p} + \nu_p  \bm x^\transp \bm x) 
      - q_{p} ( \widetilde V_{\text{c}} - \gamma_{\text{c}}) \in \Sigma_n, \label{eq:9b}\\
    & V_{p} + (\nu_p - \epsilon)\bm x^\transp \bm x \in \Sigma_n, \label{eq:9c}\\ 
    &  -\dot{V}_{p} - \epsilon \bm x^\transp \bm x + \sum_{r=1}^{L} 
      \big[ \Theta_{p,r} + z_{pr} (V_{r} - \gamma_{r}) \big] \in \Sigma_n, 
 \label{eq:9d}\\
    & y_{p} (V_{p} - \gamma_{p}) - ( \dot{V}_{p} + \epsilon \bm x^\transp \bm x) \in \Sigma_n, \label{eq:9e}\\
    & l_p (\widetilde V_{\text{c}} - \gamma_{\text{c}}) - ( \dot{\widetilde V_{\text{c}}} + \epsilon \bm x^\transp \bm x) \in \Sigma_n, \label{eq:9f}\\
    & s_{p}, q_{p}, l_p, y_{p}, z_{pr} \in \Sigma_n, \,\, \text{for} \,\, r\in\{1,\ldots,L\},  \label{eq:9g} \\
    & A_{pr} - B_{p r} \in \R_{\geq 0},    
    r \in \{1, \dots, L\} \backslash \{p\}, \label{eq:9h}\\
    & -\sum\nolimits_{r=1}^{L}  (A_{p r} - B_{p r})\in \R_{\geq 0}, \,\, \sum\nolimits_{r=1}^{L}  B_{p r}\in \R_{\geq 0}. \label{eq:9i}
\end{align}
\end{subequations}
\vspace{-0.4cm}

\noindent
Here, $\Xi = \{s_{{p}}, q_{p}, l_p, y_{p}, z_{pr},  A_{p r}, B_{p r}\}$ are the decision variables, $\Sigma_n$ denotes the set of SOS polynomials, and $\R_{\geq 0}$ denotes the set of non-negative real numbers. A small positive scalar $\epsilon $  is introduced to mitigate numerical errors.
Following the Positivstellensatz principle \cite{papachristodoulou2002construction}, constraint \eqref{eq:9b} enforces the set inclusion $\Omega_{p} \subseteq \Omega$. To guarantee that \( V_{p} \) satisfies the Lyapunov function conditions and inequality \eqref{eq.lem.timederivative}, we incorporate constraints \eqref{eq:9c}--\eqref{eq:9e}. Constraint \eqref{eq:9f} imposes that the derivative of the distributed approximation of the Lyapunov function, $\dot{\widetilde V_{\rm c}}$, is negative definite. Constraint \eqref{eq:9g} ensures the SOS optimization feasibility via the S-procedure\cite{packard2010help}, while constraints \eqref{eq:9h}--\eqref{eq:9i} enforce conditions (i)--(iii) of Theorem~\ref{thm.locallyap}.

The parameter $\nu_{p}$ is introduced to improve numerical stability of the SOS optimization. 
We seek to gradually minimize \( \nu_{p} \to 0 \) so that the partial function \( V_{p} \) converges to the composite function \( V_{\text{c}} \) within the corresponding subset. This optimization problem is formulated as
\vspace{-0.1cm}
\begin{equation}
\label{eq.minnu}
    \min_{V_{p}} \,\, \nu_{p}  \,\,\, \text{s.t.} \,\,\, \text{\eqref{eq:9b}--\eqref{eq:9i}}, \,\, V_{p}\in\Sigma_n, \,\, \gamma_{p} = \text{constant}.  
\end{equation}
\vspace{-0.4cm}

Thus, the partial function $V_{p}$ and its ROA $\Omega_{p}$ are iteratively optimized. The inner-iteration loop maximizes the level set $\Omega_{p}$ through Eq.~\eqref{eq.sosoptimization} for a fixed partial Lyapunov function $V_{p}$ (which is not specified as a decision variable). The outer-iteration loop refines the SOS polynomial $V_{p}$ through Eq.~\eqref{eq.minnu} while ensuring that $\gamma_{p}$ remains invariant. 

\begin{rem}
\label{rem.initialguess}
    To solve Eq.~\eqref{eq.sosoptimization}, an initial guess for the composite Lyapunov function $\widetilde V_{\rm c}$ is required. We initialize our optimization with a quadratic function $\widetilde V_{\text{c}} = {\bm x^{\transp}}Q {\bm x}$, where $Q$ is a symmetric positive definite matrix obtained by solving the Lyapunov equation $JQ + QJ^\transp = -I_n$ and $J$ is the Jacobian matrix evaluated at the SEP $\bm x^*$.
    Once all $V_{p}$ are computed in this first iteration, we obtain a better estimate for $\widetilde V_{\rm c}$ using the approach described in Sec.~\ref{sec.reconstructedlyap}, which is then employed in subsequent iterations to further refine $V_{p}$.
\end{rem}

\begin{rem}
\label{rem.pkl}
   For large $n$, the number of polynomial terms in $\widetilde V_{\rm c}$ significantly increases the computational burden of the SOS optimization in Eqs.~\eqref{eq.sosoptimization}--\eqref{eq.minnu}. To alleviate this issue when computing $V_{p}$, we use a low-complexity approximation of $V_{\rm c}$ that relies solely on local network information. Specifically, we project $V_{\rm c}$ onto the constrained subspace  $\mathcal P_{p} = \{\bm x \in \mathcal D : \bm x_j = 0, \forall j\notin\mathcal N_{p}\}$, where $\mathcal N_{p} = \mathcal V_{p} \cup \{j\in\{1,\ldots,N\} : i\in\mathcal V_{p}, K_{ij}\neq 0\}$ is a node subset that includes $\mathcal V_p$ and its first neighbors defined by the adjacency matrix~$K$.  For instance, for the subset $\mathcal V_{p} = \{1,6\}$ depicted in Fig.~\ref{fig.diagram}b, the neighborhood is $\mathcal N_{p} = \{1,2,3,6\}$.
\end{rem}

\subsection{Reconstruction of the Composite Lyapunov Function}
\label{sec.reconstructedlyap}

Having obtained the partial functions via the SOS optimization \eqref{eq.minnu}, we propose the following composite Lyapunov function by aggregating all partial functions:
\vspace{-0.2cm}
\begin{equation}
\label{eq.distributedlyap}
    \widetilde{V}_{\text{c}} = \sum_{p=1}^{L} w_{p} V_{p}  ,
\end{equation}
\vspace{-0.4cm}

\noindent
where $w_p$ are strictly positive weights satisfying \( \sum_{p=1}^{L} w_{p} = 1 \). Since each partial functions \( V_{p} \) satisfies the Lyapunov function conditions, and \( \bm w = (w_1,\ldots,w_L)^\transp \) is a probability vector, it follows that \( \widetilde{V}_{\text{c}} \) also meets the Lyapunov function conditions. Moreover, $\widetilde V_{\rm c}$ is an SOS polynomial since all partial functions are SOS polynomials.
Taking the time derivative of $\widetilde{V}_{\text{c}}$, it follows from Eq. \eqref{eq.lem.timederivative} that
\vspace{-0.2cm}
\begin{equation}
\dot{\widetilde{V}_{\text{c}}} \leq \sum_{p=1}^{L} w_{p} \sum_{r=1}^{L} \big[ (A_{p r} - B_{p r}) V_{r}  +  B_{p r} V_{\text{c}} \big].
\end{equation}
\vspace{-0.4cm}

\noindent
Suppose the composite Lyapunov function satisfies the exponential stability condition (i.e., $\dot{\widetilde V_{\rm c}} \leq -\lambda \widetilde V_{\rm c}$). Then, there exists a decay rate $\lambda$ so that the following inequality holds:
\begin{equation}
\label{eq.inequalityvg}
 -\lambda \sum_{p=1}^{L} w_{p} V_{p}
 \leq \sum_{p=1}^{L} w_{p} \sum_{r=1}^{L} \big[ (A_{p r} - B_{p r}) V_{r}  +  B_{p r} V_{\text{c}} \big].
\end{equation}

\noindent
By relaxing Eq.~\eqref{eq.inequalityvg} using linear matrix inequalities, we can use the interior-point method to obtain the optimal weights $\bm w$. To ensure strict negativity of the Lyapunov derivative, the solution should satisfy $0 < \epsilon \ll \lambda$, as formulated next: 

\vspace{-0.2cm}
\begin{equation}
\label{eq.interiorpoint}
\begin{aligned}
    \max_{\bm w} \,\,  \lambda \quad
    \text{s.t.} \quad & \norm{\bm w}_2 = 1, \,\,
    \bm w^\transp B\bm 1_L  \geq -\lambda, \\
    &0 \geq (A-B)^\transp \bm w \geq -\epsilon.
\end{aligned}
\end{equation}
\vspace{-0.2cm}

Our distributed approach for computing $\widetilde V_{\rm c}$ follows an iterative loop: 1) solve the SOS optimization \eqref{eq.sosoptimization}--\eqref{eq.minnu} to obtain each $V_{p}$ and $\gamma_{p}$; 2) compute the composite function $\widetilde V_{\rm c}$ and the maximum decay rate $\lambda$ by solving Eq.~\eqref{eq.interiorpoint}; and 3) repeat steps 1 and 2 until $\lambda$ converges within a specified tolerance $\epsilon$.
The final estimate of the ROA is given by
\vspace{-0.05cm}
\begin{equation}
    \widetilde\Omega = \{\bm x \in \mathcal D : \widetilde V_{\rm c} \leq \eta\},
\label{eq.estimatedROA}
\vspace{-0.1cm}
\end{equation}
\noindent
where $\eta = \sum_{p=1}^{L} w_{p} \gamma_{p}$ denotes the stability margin.

\subsection{Performance Metrics}
In nonlinear stability analysis, performance evaluation typically involves comparing ROAs obtained from time-domain simulations\textemdash considered ground truths\textemdash with those estimated using Lyapunov functions. The  simulations sample an initial condition $\bm x(0)$ and test whether the resulting trajectory converges to the equilibrium within a specified time horizon (i.e., $\lim_{t \to +\infty} \norm{\bm{x}(t) - \bm{x}^*} \leq \epsilon$) . The ground-truth ROA consists of all initial conditions satisfying this convergence criterion. We propose the following two performance metrics for the estimated ROA in Eq.~\eqref{eq.estimatedROA}.

\subsubsection{Volume accuracy}
We sample $N_1$ initial conditions from an uniform distribution, $\bm x(0)\sim\mathcal U[-a,+a]$. The volume accuracy index $I_{\rm v}$ is defined as the fraction of sampled initial conditions that are predicted to lie within $\widetilde\Omega$
(according to the Lyapunov function) relative to those that satisfy the convergence criterion via time-domain simulations.

\subsubsection{Shape dissimilarity}
To evaluate how accurately the estimated set $\widetilde\Omega$ captures the morphology of ROA, we compare the radial distances from the Lyapunov-based ROA boundary with those of the ground-truth ROA. We define the initial condition as $\bm x(0) = \bm x^* + d\bm {\rho} $, where $d$ represents the distance from the equilibrium $\bm x^*$ along a direction $\bm {\rho}$, with $\norm{\bm {\rho}}=1$. We sample $N_2$ unit vectors $\bm {\rho}$ and, for each vector, find the maximum distances $d_{\rm gt}$ and $d_{\rm est}$ at which the corresponding initial state $\bm x(0)$ remains within the ground-truth and the estimated ROAs, respectively. After finding all $N_2$ distances, we define the shape dissimilarity index $I_{\rm s}$ as the Wasserstein distance between the (mean-normalized) distributions of $d_{\rm gt}$ and $d_{\rm est}$. A smaller $I_{\rm s}$ indicates a closer morphological resemblance between the estimated and ground-truth ROAs.

\section{Numerical Results}

The implementation of our distributed method\textemdash together with the case studies presented below\textemdash is publicly available in our GitHub repository: \href{https://github.com/YimingSci/Distributed-Lya-Func}{\nolinkurl{https://github.com/YimingSci/Distributed-Lya-Func}}. For details on SOS programming, we refer to the tutorial paper \cite{packard2010help}.

\subsection{Network of Coupled van der Pol Oscillators}

\begin{figure}[t]
    \centering
    \includegraphics[width=0.8\columnwidth]{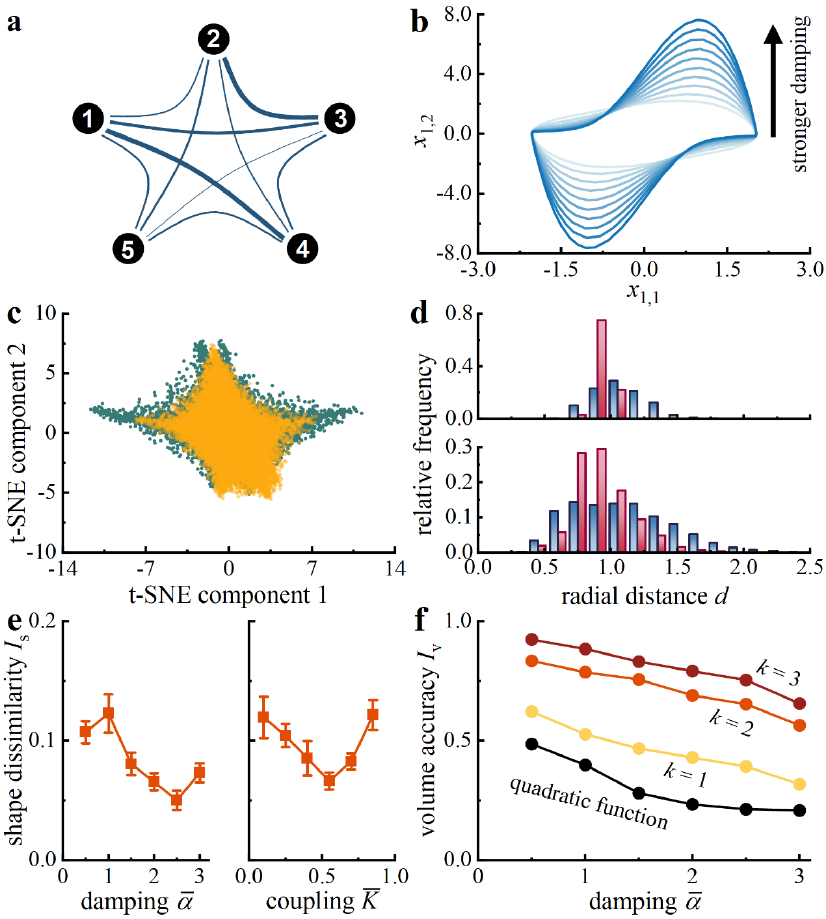}
    \caption{Optimal ROA estimation for coupled van der Pol oscillators.
    (\textbf{a})~Network of $N=5$ oscillators, where edge thickness represents coupling strength. 
    (\textbf{b})~Boundaries of the ROA in the $(x_{1,1}, x_{1,2})$-plane for increasing damping values. 
    (\textbf{c})~Estimated ROA for weak (yellow, $\overline\alpha = 0.5)$  and strong damping (green, $\overline\alpha = 3$). The t-SNE map projects all sampled initial conditions $\bm x(0)$ within the estimated ROA  onto a 2D-space.
    (\textbf{d})~Distribution of radial distances for weak (top) and strong (bottom) damping, where red and blue bars represent the estimated and ground-truth distances  $d$, respectively.
    (\textbf{e})~Shape dissimilarity $I_{\rm s}$ as a function of the average damping $\overline\alpha$ (left) and coupling strength $\overline K$ (right). The data points show averages of 10 independent parameter realizations, with error bars indicating one standard deviation. 
    (\textbf{f})~Volume accuracy $I_{\rm v}$ as a function of $\overline\alpha$ for different set cardinalities $k$. As a reference, the black line shows the result for a quadratic Lyapunov function based on the system's Jacobian matrix.
    The parameter values are  $\overline\alpha = 1$, $\sigma_{\alpha} = 0.2 \overline\alpha$, $\overline K = 0.2$,  and $\sigma_{K} = 0.2\overline K$, and we ensure that there always exists a SEP for all parameter realizations. 
    } 
    \label{fig.vanderpolshape}
\vspace{-15pt}
\end{figure}

Consider a network of $N=5$ van der Pol oscillators:
\begin{equation}
\begin{aligned}
    \dot{x}_{i,1} &= -x_{i,2},\\
    \dot{x}_{i,2} &= \alpha_i(x_{i,1}^2 - 1) x_{i,2} + x_{i,1} + \frac{1}{N} \sum_{j=1}^N K_{ij} (x_{j,1} - x_{i,1}),
\end{aligned}
\end{equation}
\noindent where $\bm x_i = (x_{i,1}, x_{i,2})^\transp$ represents the state vector of oscillator $i$, $\alpha_i$ is its damping coefficient, and  $K=(K_{ij})$ is the coupling matrix. The parameters are drawn from Gaussian distributions $\alpha_i\sim \mathcal N(\overline{\alpha},\sigma_{\alpha}^2)$ and $K_{ij}\sim \mathcal N(\overline K,\sigma_{K}^2)$. Fig.~\ref{fig.vanderpolshape}a illustrates the underlying weighted all-to-all network.
For a single oscillator, the ROA of the equilibrium $\bm x^* = 0$ is bounded by the limit cycle in the $(x_{1,1},x_{1,2})$-plane according to Poincaré–Bendixson theorem. The parameter $\alpha_i$ controls the oscillator nonlinearity: for small $\alpha_i$, the system behaves almost linearly, with a circular limit cycle, while for large $\alpha_i$ the limit cycle becomes increasingly distorted (Fig.~\ref{fig.vanderpolshape}b).

Fig.~\ref{fig.vanderpolshape}c shows the estimated $\widetilde\Omega$ provided by our composite function $\widetilde V_{\rm c}$ with set cardinality $k = 2$. We consider two scenarios: weak and strong damping. The ROA resides in the $n$-dimensional space $\mathcal S^n$, which we project (nonlinearly) onto a 2D plane using t-SNE for visualization. 
The projected ROA exhibits a complex morphology, featuring a dense cluster near the origin and structures extending outward in several directions. Note that these shapes do not imply that the estimated ROAs are non-convex, as $\widetilde V_{\rm c}$ is provably convex (Sec.~\ref{sec.reconstructedlyap}).
To evaluate the accuracy of ROA estimation, Fig.~\ref{fig.vanderpolshape}d compares the distributions of  the radial distances $d_{\rm gt}$ and $d_{\rm est}$. As $\alpha_i$ increases, the distributions become broader and more asymmetric, indicating that the high-dimensional ROA strongly deviates from a hypersphere (whose distribution collapse to a single value). 
Our results shows good morphological resemblance to the ground-truth ROA, with scores $I_{\rm s} = 0.09$ and $0.03$ for weak and strong damping, respectively.

Fig.~\ref{fig.vanderpolshape}e evaluates the performance of our approach over a wide range of parameters, showing that $I_{\rm s}$ is consistently below {0.15} on average. 
We also examine the volume accuracy of the estimated ROA in Fig.~\ref{fig.vanderpolshape}f. For $k=2$, we obtain {$I_{\rm v}\approx 75\%$} at $\overline\alpha = 0.5$, which gradually decreases to {$60\%$} as the damping increases. 
This trend is expected as, for high damping, the ground-truth ROA becomes increasingly non-convex, ultimately limiting the accuracy of our convex approach. Nevertheless, the low values of $I_{\rm s}$ at strong damping confirm that our method correctly predicts the orientation of the ROAs.
Increasing $k$ improves volume accuracy, reaching nearly 100\% for $k=3$ at small $\overline\alpha$. 
Overall, the volume accuracy of the estimated ROAs is substantially higher than the accuracy for a classical quadratic Lyapunov function (as in Remark \ref{rem.initialguess}), even for the smallest cardinality of $k=1$.

\subsection{Network of Coupled Ising Oscillators}

\begin{figure*}[t]
    \centering
    \includegraphics[width=0.94\textwidth]{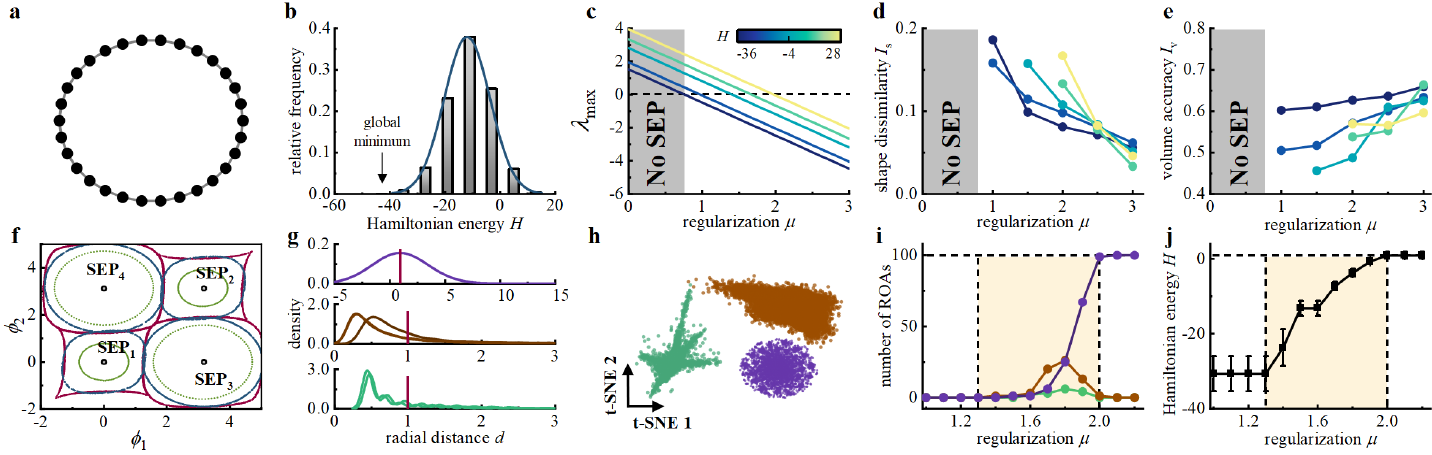} 
    \caption{Stability analysis of coupled Ising oscillators.
    (\textbf{a}) Ring network of 30 oscillators.  
    (\textbf{b}) Energy distribution across randomly sampled equilibria.
    (\textbf{c}--\textbf{e}) Largest eigenvalue $\lambda_{\rm max}$ (c), shape dissimilarity $I_{\rm s}$ (d), and volume accuracy $I_{\rm v}$ (e) as functions of $\mu$ for a subset of five SEPs, color coded by their energy.
    (\textbf{f}) ROAs of four SEPs for a subset of two Ising oscillators ($\mu=1.6$). The red, blue, and green lines indicate the boundaries of the ROAs obtained by time-domain simulations, our composite Lyapunov function, and a quadratic Lyapunov function, respectively.
    (\textbf{g}) Representative (mean-normalized) distributions of the radial distances $d$ for three types of ROA with distinct shapes.  
    (\textbf{h}) Projection of the ROA types, colored as in g.  
    (\textbf{i})~Competition between ROA types for increasing $\mu$, where the yellow shading indicates the occurrence of tentacle-like ROAs.
    (\textbf{j})~Average energy over the coexisting SEPs as a function of $\mu$, where the error bars denote one standard deviation.
    }
\label{fig.ising}
\vspace{-15pt}
\end{figure*}

Consider a network of $N$ coupled Ising oscillators \cite{cheng2024control}:
\begin{align}
\label{eq.oim}
\dot{\phi}_{i}= \sum_{j=1}^{N} K_{ij} \sin(x_j - x_i) - \mu \sin(2\phi_i),
\end{align}
\noindent
where $\phi_i$ denotes the phase of oscillator $i$, $\mu\geq 0$ is the regularization parameter, and $K\in\{-1,0\}^{N\times N}$ is a binary symmetric adjacency matrix.
This class of systems are often used to solve combinatorial optimization tasks \cite{cheng2024control}, where each equilibrium $\bm\phi^* = (\phi_1^*,\ldots,\phi_N^*)^\transp$ satisfies $\phi_i^*\in\{0,\pi\}$, $\forall i$, and is associated with a Hamiltonian energy $H = -\sum_i\sum_{j\neq i} K_{ij} \cos(\phi_i-\phi_j)$ s.t. $\sin(\phi_i) = 0$. 
Fig.~\ref{fig.ising}a,b illustrates a ring network with $N=30$ oscillators and the associated energy distribution across all possible $2^N$ equilibria, showing that only {0.04\%} equilibria correspond to the global minimum. The parameter $\mu$ controls the stability: for each equilibrium $\bm\phi^*$, the system's largest eigenvalue $\lambda_{\rm max}$  decreases linearly with $\mu$, leading to critical instability (Fig.~\ref{fig.ising}c). 

To formulate the SOS optimization \eqref{eq.sosoptimization}--\eqref{eq.minnu}, we first shift the SEP $\bm\phi^*$ to the origin and eliminate the trigonometric functions in Eq.~\eqref{eq.oim}. Using the coordinate transformation from Ref.~\cite{anghel2013algorithmic}, we define $\bm x_i = (\sin(\phi_i), 1-\cos(\phi_i))^\transp$ and introduce the constraint $x_{i,1}^2 + x_{i,2}^2 - 2x_{i,2} = 0$, $\forall i$, to the SOS formulation, ensuring it is entirely expressed in polynomial terms, with $n=2N$. Our analysis initially focuses on five equlibrium points with distinct Hamiltonian energies (Fig.~\ref{fig.ising}c--e). Note that the ROA only exists when the corresponding equilibrium is stable for sufficiently large $\mu$ (Fig.~\ref{fig.ising}c). Fig.~\ref{fig.ising}d,e shows that our method can  effectively capture the ROA morphology in large networks, yielding a consistently low $I_{\rm s}\leq 0.2$ and high $I_{\rm v}\geq 45\%$ for all SEPs (using $k=2$). 

We benchmark the method using a subsystem of two coupled Ising oscillators and four coexistent SEPs. Fig.~\ref{fig.ising}f compares the ROA obtained from: i) time-domain simulations, ii) our composite Lyapunov function, and iii) the quadratic function derived in Ref.~\cite{cheng2024control}. The results highlight that our method estimates the ROA with higher accuracy, as expected, compared to quadratic functions. 
Additionally, it also reveals that the ROAs can have distinct morphologies. 
Next, we extend our analysis to characterize the ROA morphology in the full state space. For 100 randomly selected equilibria, we identify three types of ROA shapes based on the distribution of radial distances $d$ (Fig.~\ref{fig.ising}g,h). Type I ROAs exhibit a symmetric envelope, characteristic of an $n$-dimensional ellipsoid shape. Type II ROAs show a right-skewed distribution, indicating that these ROAs extend significantly in a single direction. Type III ROAs have a multi-peak, right-skewed envelope, associated with ROAs that extend in multiple directions. 

The competition between the ROA types is shown in Fig.~\ref{fig.ising}i as the regularization parameter increases. For $\mu\leq 1.3$, all equilibria remain unstable. As $\mu$ further increases, some equilibria stabilize, with their ROAs predominantly exhibiting type II characteristics. In contrast, for $\mu\geq 2.1$, all equilibria are stable and every ROA transitions to type I. This suggests that, for sufficiently large $\mu$, the state space is approximately partitioned into regions with ellipsoid ROA boundaries (though not necessarily of uniform volume). 
Notably, type III ROAs are rare for all parameters, but they exhibit the most distinctive morphological features.
Overall, our findings suggest that the emergence of tentacle-like ROAs is associated with critical stability conditions, which occur when the average system energy is small (Fig.~\ref{fig.ising}j).

\section{Conclusion}
Our distributed approach for the construction of a composite Lyapunov function relies exclusively on local information, defined by low-dimensional subspaces specified by the state variables within a subset of nodes. The theoretical foundation is based on the proposed augmented comparison lemma, which establishes sufficient conditions for the existence of partial Lyapunov functions. These partial functions are iteratively refined using SOS optimization in order to improve the accuracy of both the volume and shape of the estimated ROAs.
We show that the resulting Lyapunov functions provide good convex approximations of the tentacle-like morphology of ROAs typical of high-dimensional network dynamics.
Future research may also explore the impact of the network structure on the ROAs and conduct a detailed analysis of the computational complexity.


\bibliographystyle{IEEEtran}
\bibliography{mybib}


\end{document}